\documentclass[12pt]{article}
\usepackage{amsmath,amsfonts,amssymb, algorithm, multirow,hyperref}
\usepackage{xcolor}
\usepackage[noend]{algpseudocode}
\algnewcommand\algorithmicforeach{\textbf{for each}}
\algdef{S}[FOR]{ForEach}[1]{\algorithmicforeach\ #1\ \algorithmicdo}
\usepackage{graphicx}
\usepackage{natbib}
\usepackage{url} 

\newcommand{\blind}{0}

\begin{document}

\def\spacingset#1{\renewcommand{\baselinestretch}%
{#1}\small\normalsize} \spacingset{1}

\newcounter{xxx}
\setcounter{xxx}{0}
\newcommand\XXX[1]{\textcolor{blue}{{\bf \em \addtocounter{xxx}{1} (\thexxx) [[#1]]}}}


\if0\blind
{
  \title{\bf Particle Data Cloning for Complex Ordinary Differential Equations 
}
  \author{Donghui Son\hspace{.2cm}\\
    Department of Statistics and Actuarial Science, Simon Fraser University \\
    and \\
    Liangliang Wang \\
    Department of Statistics and Actuarial Science, Simon Fraser University}
  \maketitle
} \fi

\if1\blind
{
  \bigskip
  \bigskip
  \bigskip
  \begin{center}
    {\LARGE\bf Title}
\end{center}
  \medskip
} \fi

\bigskip
\begin{abstract}
Ordinary differential equations (ODEs) are fundamental tools for modeling complex dynamic systems across scientific disciplines. However, parameter estimation in ODE models is challenging due to the multimodal nature of the likelihood function, which can lead to local optima and unstable inference. In this paper, we propose particle data cloning (PDC), a novel approach that enhances global optimization by leveraging data cloning and annealed sequential Monte Carlo (ASMC). PDC mitigates multimodality by refining the likelihood through data clones and progressively extracting information from the sharpened posterior. Compared to standard data cloning, PDC provides more reliable frequentist inference and demonstrates superior global optimization performance. We offer practical guidelines for efficient implementation and illustrate the method through simulation studies and an application to a prey-predator ODE model. Our implementation is available at \url{https://github.com/SONDONGHUI/PDC}.

\end{abstract}

\noindent%
\vfill

\newpage
\spacingset{1.75} 
\section{Introduction}
\label{sec:intro}


Ordinary differential equations (ODEs) are widely used to model the change rate of dynamic systems in various scientific fields, such as engineering, biology, physics, and ecology. The Lotka-Volterra model \cite[][]{lotka} is an example of such a model, which was developed by ecologists to describe the population dynamics of predators and prey within an ecosystem.
We consider a general formulation of ODEs as below: 
\begin{equation}
    \frac{d\boldsymbol{x}(t)}{dt} = \boldsymbol{g}(\boldsymbol{x}(t)|\boldsymbol{\theta}_{ode},\boldsymbol{x}(0) ), t \in [t_1, t_N],
\end{equation}
where $\boldsymbol{x}(t)=(x_1(t),\ldots,x_J(t))^{\intercal}$ denotes the ODE solutions of $J$ variables at time $t$, $\boldsymbol{g}(\cdot)$ is a prespecified function explaining the change of $\boldsymbol{x}(t)$ over time $t$, and $t_1=0$ if it is not specified otherwise. We denote unknown ODE parameters and initial conditions of ODE solutions by $\boldsymbol\theta_{ode}$ and $\boldsymbol{x}(0)$, respectively. 

Estimating the parameter vectors from real data is a topic of considerable interest for interpreting dynamic systems. However, it is highly challenging to estimate parameters in ODE models. Primarily, many ODEs are nonlinear and lack closed-form analytic solutions, complicating the parameter estimation process except in a few cases. Additionally, measurements of the systems governed by these differential equations are frequently observed with measurement errors, further making the estimation of parameters difficult. Moreover, the solutions of ODEs are often highly sensitive to the values of the parameters, resulting in a likelihood surface characterized by numerous local modes. 

In recent decades, various likelihood estimation methods have been developed to obtain maximum likelihood estimators (MLEs) of parameters in ODE models. The nonlinear least squares method \cite[][]{bard1974nonlinear, biegler1986nonlinear} is one such approach that seeks to identify optimal ODE parameter values by minimizing the discrepancy between the numerical solutions of the ODEs and the observed data. 
A two-step estimation approach utilizing regression splines was proposed by \cite{varah1982spline}. Similarly, \cite{liang2008parameter} employed a comparable two-step nonparametric regression method within the framework of measurement errors. These methods fit splines to the observations in the first step, and the ODE parameters are estimated via least squares. However, the estimated ODE parameters frequently exhibit substantial bias because the ODEs were not incorporated into the estimation process of the derivatives. To circumvent this limitation, \cite{ramsay2007parameter} proposed a generalized profiling method 
which jointly considers the smoothing of data trajectories and estimation of ODE parameters with a relatively low computational load. 
Under mild conditions, the resulting estimates are consistent and asymptotically normally distributed \cite[][]{qizhao}. \cite{jiguorobust} extended the generalized profiling method to robustly estimate parameters using measurements with outliers. A method to estimate time-varying parameters in ODEs was proposed by \cite{cao2012penalized} as well. However, obtaining globally optimal estimates for ODE parameters is not always ensured for gradient-based optimizations of frequentist methods. This difficulty arises because the likelihood surfaces of ODE models often have multiple local modes and are highly sensitive.

To tackle the challenges of maximizing the likelihood in complex ODE models, this study focuses on the data cloning (DC) method \cite[][]{Lele2007, Lelejasa}. DC leverages Markov Chain Monte Carlo (MCMC) sampling to perform frequentist inference by introducing $k$ independent data copies. This approach allows for the construction of a $k$-cloned posterior distribution, 
which is formed by raising the likelihood to the power of $k$ while maintaining the original prior distributions. As shown by 
\cite{Lelejasa}, the $k$-cloned posterior distribution converges in distribution to a multivariate normal (MVN) centered at the MLEs, effectively suppressing local maxima as $k$ increases. Therefore, the MLEs and their asymptotic variance can be estimated using samples from the $k$-cloned posterior distribution.

Data cloning has been widely applied to various models with complex and parameter-sensitive likelihoods. 
Originally introduced by \cite{Lele2007,Lelejasa} to sidestep the heavy integrations in estimating MLEs of generalized linear mixed models, DC has been extended to time series models with latent variables by \cite{JACQUIER2007615}. \cite{phylogenyanddc} employed data cloning to assess the identifiability of discrete parameters, such as tree topology in phylogenetics.  Recent work by \cite{son2023} demonstrated the utility of DC in ODE models for gene regulatory networks, highlighting that DC produces proper MLEs and consistent results regardless of prior selections.
However, their implementation relies on a simple Metropolis-Hastings (MH) algorithm \cite[][]{hastings1970}, which presents certain challenges.  Since the posterior landscape of ODE models is highly sensitive to parameter changes and often contains multiple local maxima, their MH-based DC approach is prone to converging to a local mode.  Moreover, in DC, the $k$-cloned posterior surface becomes
increasingly sharp as $k$ grows due to the exponentiated likelihood, making it difficult to generate representative samples efficiently from Markov chains. This sharpness exacerbates mixing issues and reduces acceptance rates when using a random walk (RW) MH sampler, leading to poor exploration of the posterior space. Despite these limitations, few studies have investigated ways to enhance the efficiency of DC or examined whether DC estimates can be reliably guaranteed to achieve global optimality.

Sequential Monte Carlo (SMC) methods have been widely used to overcome the obstacles of MCMC techniques \cite[][]{Neal2001, Chopin2002, DelMoral2006, smcinvitation}. In particular, a comprehensive SMC framework called SMC sampler or annealed SMC \cite[][]{DelMoral2006} has significantly increased the applicability of SMC methods beyond their traditional use in state-space models. This general algorithm gradually approaches samples representing the target by introducing annealed intermediate distributions, whereas MCMC algorithms directly generate samples from the target posterior using converged MCMC chains. An adaptive SMC sampling technique was provided by \cite{zhou2016} as a method for conducting model selection and parameter estimation. \cite{Wang2020} developed an annealed SMC (ASMC) algorithm for phylogenetics by constructing an artificial sequence of intermediate distributions. \cite{Shijia2022} applied ASMC to estimate parameters of semi-parametric nonlinear differential equations in a Bayesian sense. ASMC iterates over resampling, propagation, and reweighting steps while exploring an artificial annealed sequence of distributions. This particle-based sampler is a semi-automatic algorithm that necessitates minimal user intervention for tuning \cite[][]{Wang2020,Shijia2022}. In addition, particles can propagate mostly in parallel \cite[][]{murray2016}. As an effort to upgrade data cloning, data-cloning $\text{SMC}^{2}$ has been proposed by \cite{duan2020} to combine data cloning with $\text{SMC}^{2}$ \cite[][]{smc2chopin, FULOP2013, duan2015}. However, their approach is not suitable for obtaining MLEs of ODE models as the two layers of the SMC method are redundant to dealing with ODEs whose solutions are deterministic once the parameter values are given. 

This paper introduces a particle-based global optimizer called particle data cloning (PDC) and presents an efficient implementation of the DC algorithm to enable stable frequentist inference for ODE parameters. PDC integrates ASMC techniques into the data cloning framework, so our method takes advantage of DC while addressing its limitations. Primarily, PDC mitigates the local trapping issues often encountered in gradient-based frequentist methods by suppressing local modes through the $k$-powered likelihood of ODE models. Unlike the MCMC-based data cloning method which directly samples from the $k$-cloned posterior distribution---a challenging task when $k$ is large---PDC employs a sequence of intermediate distributions beginning with a reference distribution. This approach facilitates exploration of the posterior space even for large values of $k$, making PDC a more reliable global optimizer than the standard DC. Our method can also enhance robustness to multimodality compared to the existing data cloning and generalized profiling approach, owing to the incorporation of ASMC \cite[][]{Schweizer2012NonasymptoticEB,jasra2015, Shijia2022}. 
Furthermore, we develop a novel adaptive scheme to monitor the convergence of data cloning as $k$ increases. This scheme efficiently recycles information from the $k$-cloned posterior with smaller $k$ via reference distributions. As a result, PDC proves to be more practical and efficient to implement than the standard DC. Through extensive numerical experiments, we demonstrate the effectiveness of PDC as a global optimizer in obtaining MLEs of ODE parameters.

The rest of the paper is organized as follows. In the next section, a detailed overview of data cloning for ODE models will be presented. The particle data cloning will then be described in Section \ref{sec:gdcasmc}, along with the implementation directions for our proposal. The approach will be assessed in Sections \ref{sec:sim} and \ref{sec:real} using simulation experiments and real data analysis based on ODE models. The results demonstrate that the proposed method operates more efficiently than the original data cloning and that PDC outperforms the widely used generalized profiling method as a global optimizer. We conclude the study with a discussion in Section \ref{sec:conc}.

\section{Methodology}
\label{sec:gdcasmc}
We denote the unknown ODE parameters $\boldsymbol\theta_{ode} = (\theta_{ode_1},\ldots,\theta_{ode_P})^{\intercal}$, where $P$ is the number of ODE parameters. The initial conditions of ODE solutions are denoted by $\boldsymbol{x}(0)=(x_1(0),\ldots,x_J(0))^{\intercal}$, where $J$ is the number of ODE components. 
Since only a subset of the $J$ ODE components is observed in many cases, we let $\mathcal{J}_{obs} \subseteq \{1,\ldots,J\}$ denote this subset. 
Throughout the studies in this article, $\boldsymbol{y}_{j} = (y_{1{j}},\ldots,y_{N{j}})^{\intercal}$ denotes the data for the $j$th ODE variable where $j \in \mathcal{J}_{obs}$, and $\boldsymbol{y}$ is the aggregation of $\boldsymbol{y}_j$  for all $j$  . The $i$th observation of $\boldsymbol{y}_j$ is observed with an additive measurement error according to
\begin{equation}\label{ode_normal2}
    y_{ij}=x_j(t_{ij}) + \epsilon_{ij}, i=1,\ldots,N,
\end{equation} 
where $\epsilon_{ij}$ is assumed to follow $N(0,\sigma_j^2)$ for all time.
We assume for simplicity that the number of observations for each measured ODE variable is the same and denoted as $N$, but this can be also extended to accommodate irregular observation models by introducing the notation $N_j$.

Under the Gaussian error assumption (Equation \ref{ode_normal2}), the joint likelihood function of $\boldsymbol{\theta}=(\boldsymbol\theta_{ode}^{\intercal}, {\boldsymbol{x}(0)^{\intercal},{\boldsymbol\sigma^2}^{\intercal})}^{\intercal}$ is given by
\begin{equation}\label{ode_like}
    p(\boldsymbol{y}|\boldsymbol\theta) \propto \prod_{i=1}^{N}\prod_{j \in \mathcal{J}_{obs}}^{}(\sigma^2_{j})^{-\frac{1}{2}} \exp\left\{{-\frac{(y_{ij}-x_j(t_{ij}|\boldsymbol\theta_{ode},\boldsymbol{x}(0)))^2}{2\sigma^2_j}}\right\},
\end{equation} 
where $i=1,\ldots,N, j \in \mathcal{J}_{obs},$ and $\boldsymbol\sigma^2 = \{ \sigma^2_{j} \}_{j \in \mathcal{J}_{obs}}$. 
As an example, the log-likelihood surface for ODE parameters under the setting of the model in Section \ref{sec:sim} has multiple isolated modes, and it is very sensitive to different values of parameters (see Figure 1 (b) of \cite{Shijia2022}).

\subsection{MLEs of ODE Models with Data Cloning}
This subsection provides a brief overview of how to obtain MLEs of parameters in ODE models using data cloning \cite[][]{Lele2007,Lelejasa}. We let $\boldsymbol{y}^{(k)}=(\boldsymbol{y}^{\intercal},\boldsymbol{y}^{\intercal},\ldots,\boldsymbol{y}^{\intercal})^{\intercal}$ denote the cloned data vector which results in the $k$-powered likelihood function $p(\boldsymbol{y}^{(k)}|\boldsymbol{\theta})=p(\boldsymbol{y}|\boldsymbol{\theta})^{k}$ under the independence assumption of data cloning. Based on the likelihood with $k$ clones, the $k$-cloned posterior distribution $\pi^{(k)}(\boldsymbol{\theta}|\boldsymbol{y})$ is defined as follows:
\begin{align}\label{kclonedpost}
\pi^{(k)}(\boldsymbol{\theta}|\boldsymbol{y}) \propto p(\boldsymbol{y}^{(k)}|\boldsymbol{\theta}) p_0(\boldsymbol{\theta})=p(\boldsymbol{y}|\boldsymbol{\theta})^{k} p_0(\boldsymbol{\theta}_{ode}) p_0(\boldsymbol{x}(0))p_0(\boldsymbol\sigma^2), 
\end{align}
where $p_0(\cdot)$ denotes a prior distribution for the parameters. 
The Metropolis-Hastings algorithm, one of the standard MCMC techniques, has been applied to obtain samples from this $k$-cloned posterior distribution \cite[][]{Lele2007, Lelejasa}.

Once we aggregate proper $L$ MCMC samples after discarding the first half of the iterations as the burn-in period, data cloning enables us to derive the MLE vector $\hat{\boldsymbol{\theta}}$ and its asymptotic variance-covariance matrix $I^{-1} (\hat{\boldsymbol{\theta}})$. As shown by \cite{Lelejasa}, the $k$-cloned posterior converges in distribution to a multivariate normal (MVN) distribution for a sufficiently large $k$ as below:
\begin{equation}\label{theorydc}
    \pi^{(k)}(\boldsymbol{\theta}|\boldsymbol{y}) \xrightarrow[]{d} MVN(\hat{\boldsymbol{\theta}}, k^{-1} I^{-1} (\hat{\boldsymbol{\theta}})),
\end{equation}
where $I(\hat{\boldsymbol{\theta}})$ is the Fisher information matrix for the original likelihood.
This result implies that $\pi^{(k)}(\boldsymbol{\theta}|\boldsymbol{y})$ nearly degenerates at $\hat{\boldsymbol{\theta}}$ and that $I^{-1} (\hat{\boldsymbol{\theta}})$ can be recovered by multiplying the variance component of Equation \ref{theorydc} by $k$. Moreover, data cloning mitigates issues related to prior distribution sensitivity, which can be problematic with small sample sizes in Bayesian estimations \cite[for details, see][]{Lele2007, Lelejasa,duan2020, son2023}.
Algorithm \ref{dcalgo} describes the data cloning algorithm with MH steps.
\begin{algorithm}[t]
	\caption{Data Cloning with the MH algorithm}
        \label{dcalgo}
	\begin{algorithmic}[1]
         \State \textbf{Inputs:} (a) Prior distribution over model parameters $p_0(\boldsymbol{\theta})$; (b) Likelihood function $p(\boldsymbol{y}|\boldsymbol{\theta})$; (c) The number of clones $k$.
         \State \textbf{Outputs:} Posterior approximation $\hat{\pi}^{(k)}(\boldsymbol{\theta}|\boldsymbol{y})=\sum^{2L}_{l=L+1} L^{-1}\delta_{\boldsymbol{\theta}^{(l)}}(\boldsymbol{\theta})$; MLEs $\hat{\boldsymbol{\theta}}$ and asymptotic variance $I^{-1} (\hat{\boldsymbol{\theta}})$.
         \State Start with an initial sample $\boldsymbol{\theta}^{(0)}$.
		\For {$l=1,2,\ldots,2L$}
		\State Propose a move to a new location, $\boldsymbol{\theta}^{new}$, by drawing a sample from a proposal distribution, $q(\boldsymbol{\theta}^{(l-1)}, \boldsymbol{\theta}^{new})$.
        \State Accept the move (i.e., set $\boldsymbol{\theta}^{(l)}=\boldsymbol{\theta}^{new}$) with probability,
        \begin{equation}
            \min\left\{1, \frac{p(\boldsymbol{y}|\boldsymbol{\theta}^{new})^{k} p_0(\boldsymbol{\theta}^{new})}{p(\boldsymbol{y}|\boldsymbol{\theta}^{(l-1)})^{k} p_0(\boldsymbol{\theta}^{(l-1)})} 
            \cdot \frac{q(\boldsymbol{\boldsymbol{\theta}^{new},\theta}^{(l-1)})}{q(\boldsymbol{\theta}^{(l-1)}, \boldsymbol{\theta}^{new})} \right\} \nonumber.
        \end{equation}
		\EndFor
  \State Compute posterior sample mean vector and variance to obtain MLEs and its asymptotic variance-covariance matrix,
  \begin{equation}\label{dcoutput}
      \hat{\boldsymbol{\theta}}=\frac{1}{L} \sum^{2L}_{l=L+1} \boldsymbol{\theta}^{(l)}, ~I^{-1} (\hat{\boldsymbol{\theta}}) = \frac{k}{L-1} \sum^{2L}_{l=L+1}(\boldsymbol{\theta}^{(l)}-\hat{\boldsymbol{\theta}})(\boldsymbol{\theta}^{(l)}-\hat{\boldsymbol{\theta}})^{\intercal}.
  \end{equation}
\end{algorithmic} 
\end{algorithm}

\subsection{Particle Data Cloning}\label{asmc}
MCMC-based data cloning in Algorithm \ref{dcalgo} can produce the MLEs of ODE parameters and their uncertainty in simple cases. However, this algorithm may become inefficient for a complex ODE model due to the practical hurdles of MCMC methods. First, the original data cloning using a simple MCMC may become trapped in specific modes, as the posterior distribution of a complex ODE model is highly sensitive to parameter values and initial states, exhibiting multimodality in high-dimensional spaces. Second, the final inferences from the MCMC chains targeting the $k$-cloned posterior may become unreliable since the posterior structure with a large $k$ could have an excessively sharpened shape due to Equation \ref{theorydc}. Thus, we face a paradoxical situation where we need to explore the posterior space with sufficiently large $k$ to obtain an accurate global optimum while this task becomes more challenging as $k$ increases.

To address the challenges in MCMC-based data cloning, we propose particle data cloning (PDC) to gradually approximate $\pi^{(k)}(\boldsymbol{\theta}|\boldsymbol{y})$ by introducing a sequence of intermediate distributions $\{\pi_r(\boldsymbol{\theta})\}_{r=0}^{R}$, unlike MCMC-based data cloning which directly samples from the target. This sequence of intermediate distribution $\{\pi_r(\boldsymbol{\theta})\}_{r=0}^{R}$ starts from a reference distribution $\Tilde{\pi}(\boldsymbol{\theta})$ \cite[][]{Fan2011}
 and ends with the target density $\pi_R(\boldsymbol{\theta}) = \pi^{(k)}(\boldsymbol{\theta}|\boldsymbol{y})$. 
  PDC updates an approximation to these densities iteratively.

We use annealing \cite[][]{Neal2001, DelMoral2006, Wang2020} to create a sequence of intermediate distributions that allow our samplers to explore the flattened posterior space with flexibility. The $r$-th intermediate distribution for ODE models can be defined as below 
 \begin{align}\label{kcloneintermed}
    \pi_r(\boldsymbol{\theta})\propto \gamma_r(\boldsymbol{\theta}) &= [p(\boldsymbol{y}|\boldsymbol{\theta})^{k} p_0(\boldsymbol{\theta})]^{\phi_r} \Tilde{\pi}(\boldsymbol{\theta})^{1-\phi_r}, 
\end{align}
where the sequence of annealing parameters $\{\phi_r\}_{r=0}^{R}$ satisfies $0=\phi_0 < \phi_1 < \cdots <\phi_{R-1} < \phi_{R} =1$. These parameters can be either pre-specified or adaptively determined during the implementation of annealed SMC \citep{zhou2016,Shijia2022}.  It is also possible to choose other distribution sequences \cite[see, e.g][]{Chopin2002, DelMoral2006,chopin2013}. 

For data cloning analysis, appropriate priors should be assigned to unknown parameters $\boldsymbol{\theta}_{ode}, \boldsymbol{x}(0), \sigma_j^2, j \in \mathcal{J}_{obs}$. We specify them as follows: 
\begin{equation}\label{prior}
        \boldsymbol{\theta}_{ode} \sim MVN_P\left(\boldsymbol{\mu},\Sigma_{ode}\right);
    \boldsymbol{x}(0)\sim MVN_J\left(\boldsymbol{\tau},\Sigma_0\right); \sigma_j^2 \sim IG(a,b), \nonumber
\end{equation} 
where $\boldsymbol{\mu}=(\mu_1,\ldots,\mu_P)^{\intercal}$, and $\Sigma_{ode}=\text{diag}\left(\sigma^2_{{ode_1}},\ldots,\sigma^2_{{ode_P}}\right)$ are the hyperparamters in the prior distribution of $P$ dimensional ODE parameters $\boldsymbol{\theta}_{ode}$. For the $J$ dimensional initial conditions $\boldsymbol{x}(0)$, we use $\boldsymbol{\tau}=(\tau_1,\ldots,\tau_J)^{\intercal}$ and $\Sigma_{0}=\text{diag}\left(\sigma^2_{01},\ldots,\sigma^2_{0J}\right)$ as the hyperparameters. Also, $a$ and $b$ are the hyperparameters in Inverse Gamma prior for the variance parameters.

PDC involves important steps of reweighting, resampling, and particle movement because it combines data cloning with annealed SMC. At each step $r$, the approximations to each density are represented in terms of a collection of $M$ particles and their respective normalized weights,  $\{\boldsymbol{\theta}_r^{(m)}, W_r^{(m)}\}_{m=1}^{M} \sim \pi_r(\boldsymbol{\theta})$. 
 Suppose that a set of weighted $M$ particles $\{\boldsymbol{\theta}_{r-1}^{(m)}, W_{r-1}^{(m)}\}_{m=1}^{M}$ approximating $\pi_{r-1}(\boldsymbol{\theta})$ is available, and $\{W_{r-1}^{(m)}\}_{m=1}^{M}$ are the corresponding normalized weights. The $(r-1)$th particles should be updated to the counterparts of the next step, $\{\boldsymbol{\theta}_{r}^{(m)}, w_{r}^{(m)}\}_{m=1}^{M}\sim \pi_{r}(\boldsymbol{\theta})$. We need two sequences of Markov kernels, forward and backward kernels. The forward Markov kernel $K_{r}(\boldsymbol{\theta}_{r-1}^{(m)}, \boldsymbol{\theta}_{r}^{(m)})$ is designed to propose new particles based on a certain MCMC kernel in practice \cite[][]{Chopin2002, DelMoral2006,JASRA2011, fearnhead2013, zhou2016}. We illustrate in detail how to construct the forward kernel in subsection \ref{markovk}. For the backward kernel, we use $B_{r-1}(\boldsymbol{\theta}_{r}^{(m)},\boldsymbol{\theta}_{r-1}^{(m)})=\pi_{r}(\boldsymbol{\theta}_{r-1}^{(m)}) 
K_{r}(\boldsymbol{\theta}_{r-1}^{(m)}, \boldsymbol{\theta}_{r}^{(m)}) / \pi_{r}(\boldsymbol{\theta}_{r}^{(m)})$ to simplify the weight updating process \cite[][]{DelMoral2006, Wang2020}. Thus, the new weights can be obtained using the previous particles and the two kernels according to
\begin{align}\label{weightfuncdc_incremental} 
    \tilde{w}_r^{(m)} &= [p(\boldsymbol{y}|\boldsymbol{\theta}_{r-1}^{(m)})^k p_0(\boldsymbol{\theta})/\Tilde{\pi}(\boldsymbol{\theta})]^{(\phi_r-\phi_{r-1})}, \\ \label{weightfuncdc_unnormalized} 
    w_r^{(m)} &= w_{r-1}^{(m)} \cdot \tilde{w}_r^{(m)} ,
\end{align}
and we can calculate the corresponding normalized weights by $W_{r}^{(m)} = w_{r}^{(m)}/(\sum^{M}_{m=1} w_{r}^{(m)})$.

Starting from initializing particles with equal weights $\{\boldsymbol{\theta}_0^{(m)}, \frac{1}{M}\}_{m=1}^{M}$, PDC aims to obtain $\{\boldsymbol{\theta}_R^{(m)}, W^{(m)}_R\}_{m=1}^{M}$, which represents the $k$-cloned posterior distributions with a fixed $k$.
\begin{equation}
    \hat{\pi}_R(\boldsymbol{\theta}) = \hat{\pi}^{(k)}(\boldsymbol{\theta}|\boldsymbol{y})=\sum_{m=1}^{M} W_R^{(m)}\cdot\delta_{\boldsymbol{\theta}_R^{(m)}}(\boldsymbol{\theta}) \nonumber
\end{equation}
where $\delta_{\boldsymbol{\theta}_R^{(m)}}(\boldsymbol{\theta})$ denotes a point mass on $\boldsymbol{\theta} = \boldsymbol{\theta}_R^{(m)}$. Once the weighted particles are collected in the last step $R$, we can easily obtain MLEs and their asymptotic variances using weighted mean and covariance as follows,
\begin{equation}\label{smcsummary}
\hat{\boldsymbol{\theta}}= \sum^{M}_{m=1}W_R^{(m)} \boldsymbol{\theta}_R^{(m)}, ~I^{-1} (\hat{\boldsymbol{\theta}}) = k\sum^{M}_{m=1}W_R^{(m)}(\boldsymbol{\theta}_R^{(m)}-\hat{\boldsymbol{\theta}})(\boldsymbol{\theta}_R^{(m)}-\hat{\boldsymbol{\theta}})^{\intercal}.
\end{equation}
 To alleviate the particle degeneracy problem, we conduct a multinomial resampling method based on the normalized particle weights to prune particles with small weights. 
We use the relative effective sample size (rESS) as a criterion to trigger resampling. PDC determines resampling if $\text{rESS}_{r}^{(M)} = \text{ESS}_{r}^{(M)}/M < \zeta$, where the effective sample size \cite[ESS, see, e.g][]{kong1994, liusmc1998} is defined as $\text{ESS}_r^{(M)} =1/\sum_{m=1}^{M}(W_r^{(m)})^2$ at each step $r$. We set $\zeta=0.5$ throughout this paper. After resampling, an equally weighted particle set is obtained. Other resampling schemes can also be considered, such as stratified resampling \cite[][]{Kitagawa, Hol2006}, residual resampling \cite[][]{doucappe}, or systematic resampling \cite[][]{carpenter}.

We implement an adaptive annealing parameter scheme based on the relative conditional effective sample size (rCESS) \citep{zhou2016}. Given a user-specified  threshold $\varepsilon_{\text{rCESS}}\in (0,1]$, typically set close to 1, the annealing parameter at the $r$th iteration is determined by solving the following equation: 
\begin{equation}\label{rCESS}
f(\phi_r) =  \text{rCESS}_r\left(W_{r-1}^{(\cdot)}, \tilde{w}_{r}^{(\cdot)}\right) = \frac{\left(\sum^M_{m=1} W_{r-1}^{(m)} \tilde{w}_{r}^{(m)}\right)^2}{\sum^{M}_{m = 1} W_{r-1}^{(m)}\left(\tilde{w}_{r}^{(m)}\right)^2} = \varepsilon_{\text{rCESS}}.
\end{equation}
Since it is not possible to obtain a closed-form solution from Equation \ref{rCESS}, we use a bisection method that searches the interval $(\phi_{r-1},1]$ by using $f(\phi_{r-1})-\varepsilon_{\text{rCESS}}>0, f(1)-\varepsilon_{\text{rCESS}}<0$ (we set $\phi_r=1$ if $f(1)\geq \varepsilon_{\text{rCESS}}$). This formulation ensures that the annealing parameter $\phi_r$ is adaptively chosen at each iteration to maintain the specified rCESS threshold.

 \subsection{Forward Markov Kernels}\label{markovk}
 A forward Markov kernel $K_{r}(\boldsymbol{\theta}_{r-1}^{(m)}, \boldsymbol{\theta}_{r}^{(m)})$ is required to propagate the previous particles with new weights. Among the options described in \cite{DelMoral2006}, we use one of the MCMC kernels as our forward kernel. Since the full conditional posteriors cannot be derived excluding the variance parameter $\sigma_j^2$, we generate new particles $\{\boldsymbol{\theta}_{r}^{(m)} \sim K_{r}(\boldsymbol{\theta}_{r-1}^{(m)}, \cdot) \}_{m=1}^{M}$ via the MH-Gibbs kernel.
 
 We mainly utilize the adaptive MH algorithm \cite[][]{haario2001adaptive,adaptivemcmc}  to design MH-Gibbs moves for the forward kernel. The MH-Gibbs move is $\pi_{r}$-invariant, which targets the $r$th intermediate $k$-cloned posterior in Equation \ref{kcloneintermed}. The full conditional posterior distribution of $\sigma_j^2$ is defined as follows for all $j \in \mathcal{J}_{obs}$:
 \begin{equation}
     \sigma^2_j | - \sim IG \left( a+\frac{Nk\phi_r}{2},
     b+\frac{k\phi_r}{2} \sum^N_{i=1} (y_{ij}-x_j(t_{ij}|\boldsymbol\theta_{ode},\boldsymbol{x}(0)))^2\right),\nonumber
 \end{equation}
but the full conditional posterior distribution for $\boldsymbol{\theta}_{ode}$ and $\boldsymbol{x}(0)$ is known only up to the normalizing constant without a closed form,
\begin{align}
    &\pi_r(\boldsymbol\theta_{ode},\boldsymbol{x}(0)|-) \propto \nonumber\\ 
    &\exp\left[ -k\phi_r \left\{\sum^N_{i=1}\sum^{}_{j \in \mathcal{J}_{obs}} \frac{(y_{ij}-x_j(t_{ij}|\boldsymbol\theta_{ode},\boldsymbol{x}(0)))^2}{2\sigma^2_j} \right\} 
    -\sum^{J}_{j=1}\frac{(x_j(0)-\tau_j)^2}{2\sigma^2_{0j}} -\sum^P_{p=1}\frac{(\theta_{ode_p}-\mu_p)^2}{2\sigma^2_{{ode_p}}} \right] \nonumber.
\end{align}

When propagating particles for these two parameters, candidate particles are drawn from a normal mixture distribution, 
\begin{equation}
 0.95\cdot N\left(\boldsymbol{\theta}_{r-1}^{(m)}, \frac{(2.38)^2}{(P+|\mathcal{J}_{obs}|)}\hat{\Sigma}_{r-1}\right)+0.05 \cdot N\left(\boldsymbol{\theta}_{r-1}^{(m)}, \frac{(0.1)^2}{(P+|\mathcal{J}_{obs}|)} I_{(P+|\mathcal{J}_{obs}|)}\right),   
\end{equation}
where $|\mathcal{J}_{obs}|$ denotes the cardinality of the observable set of ODE variables, and $\hat{\Sigma}_{r-1}$ is the empirical weighted covariance estimate based on the $M$ particles from the $(r-1)$th intermediate distribution. The proposed particles are accepted according to the usual acceptance rule of MH algorithm. We describe the whole PDC procedure with these details in Algorithm \ref{dcasmc}. 

The forward kernel in PDC is closely related to the MCMC kernel employed in the existing data cloning approach described in Algorithm \ref{dcalgo}. However, the key difference lies in how these kernels target the $k$-cloned posterior distribution. In standard DC, regardless of the proposal distribution, the MCMC kernel aims to sample from the target $k$-cloned posterior, maintaining $\pi$-invariance. In contrast, PDC employs a stepwise refinement strategy for the intermediate $k$-cloned posteriors; the $\pi_r$-invariant PDC kernels gradually transform weighted particles from a reference distribution to the $k$-cloned posterior. PDC can be implemented using two candidate forward Markov kernels: (1) $\pi_r$-invariant RWMH-Gibbs moves, which rely on a random walk Metropolis-Hastings update within a Gibbs sampling framework, and (2) $\pi_r$-invariant adaptive MH-Gibbs moves, which incorporate adaptive proposal distributions to improve efficiency. The relationship between these kernels and those in DC is straightforward: the standard DC kernels can be derived by simply removing the annealing parameter $\phi_r$ from the corresponding PDC kernel.  In Section \ref{sec:real}, we compare the performance of PDC and DC under these different kernel choices.

\subsection{Adaptive PDC for a Sequence of Clone Numbers}
\begin{figure}[htp]
    \centering
    \includegraphics[width=15cm]{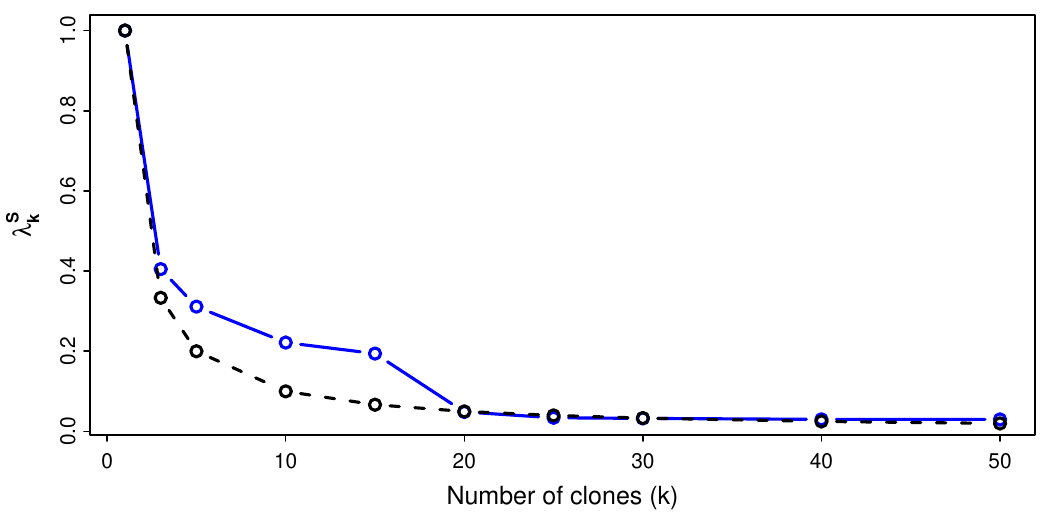}
    \caption{Data cloning convergence diagnostics for the prey-predator ODE model. The function of standardized eigenvalues ($\lambda_k^S$) converges to zero at the expected rate ($1/k$). The solid line is from particle data cloning, and the dashed line is $1/k$.}
    \label{fig:ppeigen}
\end{figure}
According to the theory of data cloning, a global maximum of a likelihood function is achievable as the number of clones $k$ becomes nearly infinite, but $k$ is finite in reality. There is a possibility that the process may fall into a local maximum with an inadequate number of clones. To determine if $k$ is sufficient, additional efforts must be made to assess whether the $k$-cloned posterior distribution is nearly degenerate after a sufficiently large $k$.

\cite{Lelejasa} suggests a graphical way to select a proper size of $k$ by investigating the behavior of the $k$-cloned posterior variance's largest eigenvalue $\lambda_k^{max}$. Here, we denote a standardized largest eigenvalue as $\lambda_k^S = \lambda_k^{max} / \lambda_1^{max}$, where $\lambda_1^{max}$ is the largest eigenvalue of the posterior variance without cloning. From the fact that $\lambda_k^{max}$ converges to zero at the same rate as $k^{-1}$, this plot of $\lambda_k^S$ against $k$ can be compared to $k^{-1}$.
Therefore, the parameters are estimable when $\lambda_k^S$ is stabilized along with $k^{-1}$, and we can select an adequate number of clones heuristically such that $\lambda_k^S$ remains below a predetermined threshold (see Figure \ref{fig:ppeigen}).

 The data cloning procedure involves incrementally increasing $k$ to identify the appropriate stopping point by monitoring the convergence of eigenvalues.
However, initiating the process from the prior distribution for each value of $k$ is computationally inefficient, particularly when generating particles from the posterior distribution for larger values of $k$. To overcome this inefficiency, our adaptive initialization scheme for PDC adjusts the reference distribution $\Tilde{\pi}(\boldsymbol{\theta})$, which enables the reuse of particles obtained from $k$-cloned posterior with smaller values of $k$.

 For the case of $k=1$, the particles are initialized directly from the prior distribution, as the available information is inherently limited. However, when weighted particles are available from the PDC results with a clone number  $k_1 > 1$, they provide a more informed basis for subsequent computations.
 Suppose that weighted particles are available from the PDC results with $k_1 > 1$, i.e., $\{\boldsymbol{\theta}_{R_{k_1}}^{(m)}\}_{m=1}^{M} \sim \hat{\pi}^{(k_1)}(\boldsymbol{\theta}|\boldsymbol{y})$, and $R_{k_1}$ denotes the last annealing index. The next objective is to generate particles from $\pi^{(k_2)}(\boldsymbol{\theta}|\boldsymbol{y})$ with a clone number $k_2 > k_1$. We propose to construct the reference distribution for $\boldsymbol{\theta}$ based on the previous PDC results: 
 \begin{equation}\label{ref}
     \boldsymbol{\theta} \sim MVN \left( \hat{\boldsymbol{\theta}}_{R_{k_1}}, \hat{\Sigma}_{R_{k_1}} \right), 
 \end{equation}
where $\hat{\boldsymbol{\theta}}_{R_{k_1}}$ is the weighted mean of the previous particles and $\hat{\Sigma}_{R_{k_1}}$ is the estimated weighted covariance matrix from the previous PDC computation. For the new implementation of PDC with $k_2$, we initialize the particles from the new reference distribution in Equation \ref{ref}.

\begin{algorithm}
	\caption{Particle Data Cloning (PDC)}
        \label{dcasmc}
	\begin{algorithmic}[1]
         \State \textbf{Inputs:} (a) Prior distribution over model parameters $p_0(\boldsymbol{\theta})$; (b) Likelihood function $p(\boldsymbol{y}|\boldsymbol{\theta})$; 
         (c) The number of clones $k$; (d) rCESS threshold $\varepsilon_{\text{rCESS}}$; (e) The resampling threshold $\zeta$.
         \State \textbf{Outputs:} Posterior approximation $\hat{\pi}_R(\boldsymbol{\theta})$; MLEs $\hat{\boldsymbol{\theta}}$ and asymptotic variance $I^{-1} (\hat{\boldsymbol{\theta}})$.
         \State Initialize the iteration index and the annealing parameter: $r \xleftarrow{} 0$, $\phi_0 \xleftarrow{} 0$.
         \ForEach {$m = 1,2,\ldots,M$}
          \State Sample initial particles:
           $\boldsymbol{\theta}_0^{(m)} \sim p_0(\cdot)$  if $k=1$, o.w. from $\Tilde{\pi}_0(\cdot)$ via Equation \ref{ref}.
        \State Initialize particle weights: $w_0^{(m)}\xleftarrow{}1, W_0^{(m)}\xleftarrow{}1/M$. 
        \EndFor
    \While{$\phi_r < 1$}
        \State $r \gets r + 1$
        \State Determine the next annealing parameter $\phi_r \in (\phi_{r-1},1]$ by solving $f(\phi_r) =  \text{rCESS}_r\left(W_{r-1}^{(\cdot)}, \tilde{w}_{r}^{(\cdot)}\right) = \varepsilon_{\text{rCESS}}$ in Equation \ref{rCESS} via a bisection method.
         
            \ForEach {$m = 1,2,\ldots,M$}
            \State Compute the incremental particle weights $\tilde{w}_{r}^{(\cdot)}$         
        using Equations \ref{weightfuncdc_incremental}.
            \State Reweight the particle weights for $\boldsymbol{\theta}_r^{(m)}$ via Equation \ref{weightfuncdc_unnormalized}.
            
            \State Normalize the particle weights: $W_r^{(m)} = w_r^{(m)}/\sum_{m=1}^{M}w_r^{(m)}$.

            \State Propagate particles with one $\pi_r$-invariant MH-Gibbs move: $\boldsymbol{\theta}_{r}^{(m)} \sim K_{r}(\boldsymbol{\theta}_{r-1}^{(m)}, \cdot)$.
            \EndFor
        \If{$\phi_r=1$}
               \State Break and return the current particles and weights $\{\boldsymbol{\theta}_r^{(m)}, W_r^{(m)}\}_{m=1}^M$.
            \Else
                 \If{$\text{rESS}_r < \zeta$}
                     \State Resample the particles.
                     \State Reset the corresponding weights: $\{w_r^{(m)}=1, W_r^{(m)}=1/M \}_{m=1}^{M}$.
                 \Else
                 \State Take weighted particles without resampling.
                 \EndIf
        \EndIf
        \EndWhile
\State Compute $\hat{\boldsymbol{\theta}}$  and $I^{-1} (\hat{\boldsymbol{\theta}})$ according to Equation \ref{smcsummary}.

	\end{algorithmic} 
\end{algorithm}

\section{Simulation Studies}\label{sec:sim}
In this section, we conduct simulation studies to verify the effectiveness of our proposed particle data cloning (PDC) compared to the existing data cloning with an MH sampler. To ensure a fair comparison, PDC is implemented using the same random walk MH proposal as the existing data cloning method. Our experiments focus on nonlinear ODEs, where the likelihood surface is highly sensitive and irregular. To assess robustness, we consider two cases: one with a unique set of true parameters and another with multiple true parameter values. In the first case, we demonstrate that our proposed data cloning method outperforms the original data cloning in terms of coverage probabilities while avoiding local trapping. In the second case, we apply PDC to nonlinear ODEs with two true parameter values to evaluate its ability to capture multiple modes more effectively than data cloning with an MH sampler. All simulations are performed using the R package \textit{deSolve} \cite[][]{2010desolve} to generate data from differential equations.

\subsection{Scenario 1: A Unique True Parameter}
We first consider a nonlinear ODE system with unique true parameters. As shown in Figure 1(b) of \cite{Wang2020}, the log-likelihood surface for $\boldsymbol{\theta} = (\theta_1, \theta_2)^{\intercal}$ in Equation \ref{5.1eq} exhibits a multimodal structure, posing a risk of local model trapping when using naive 
optimization methods. The ODE system used for generating data is
\begin{align}\label{5.1eq}
    \frac{dx_1(t)}{dt} &= \frac{72}{(36+x_2(t))} -\theta_1, \nonumber\\
    \frac{dx_2(t)}{dt} &= \theta_2 x_1(t) - 1,
\end{align}
where $\boldsymbol{\theta}=(2,1)^{\intercal}$ and initial conditions $x_1(0)=7$ and $x_2(0) =-10$. Assuming normality as in Equation \ref{ode_normal2}, we simulate 121 observations within $[0,60]$ with variance $\sigma_1 = 1$ and $\sigma_2=3$. This data generation process is repeated 50 times to calculate coverage probabilities for both our proposed method and the standard data cloning. For data cloning analysis, we specify the following prior distributions for $\theta_1, \theta_2, x_1(0), x_2(0), \sigma_1^2, \sigma_2^2$:
\begin{equation}\label{sec5prior}
        \theta_1,\theta_2 \sim N(5,5^2);~
    x_1(0), x_2(0) \sim N(2,4^2); ~\sigma_1^2, \sigma_2^2 \sim IG(1,1). \nonumber
\end{equation}
We solve the ODEs using the ``\textit{lsoda}" solver in \textit{deSolve}. In PDC, we set the rCESS threshold $\varepsilon_{\text{rCESS}} =0.999$ and the resampling threshold $\zeta=0.5$, using $M=500$ particles. For the standard data cloning (DC) method, we run 300,000 iterations, approximately matching $M\cdot R$ in PDC. 
 As described in Section \ref{sec:gdcasmc}, $\sigma_1^2$ and $\sigma_2^2$ can be sampled from their closed-form inverse gamma posterior distributions. Since closed-form updates are unavailable for ODE parameters and initial values, we use an MH step for their sampling. 

\begin{figure}[htp]
    \centering
    \includegraphics[width=15cm, height=5.5cm]{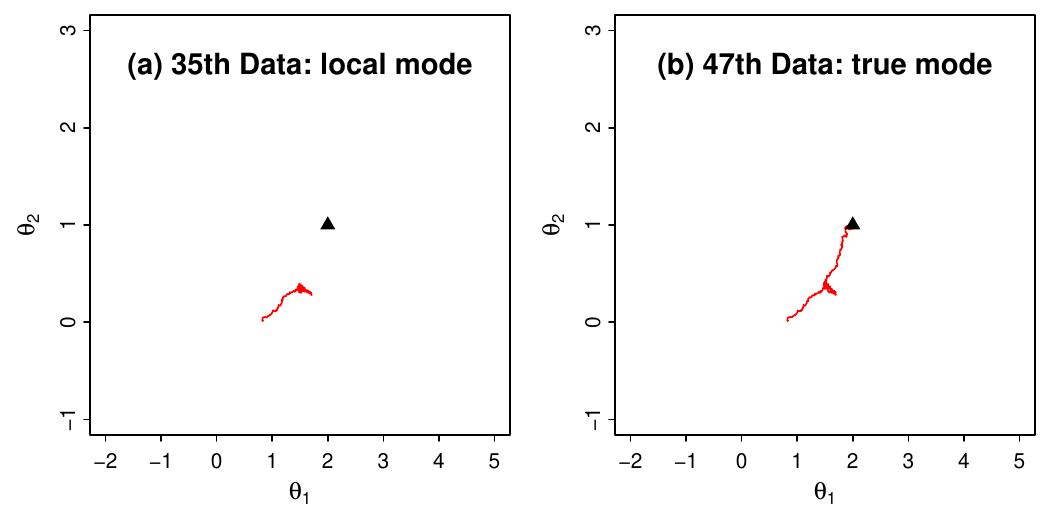}
    \caption{(a) and (b): Examples when the DC algorithm falls into a local mode (35th data) and true mode (47th data) respectively. The triangular dots indicate true parameter values of $\theta_1=2$ and $\theta_2=1$. Trace plots for $\boldsymbol{\theta}$ are presented with lines.}
    \label{fig:local}
\end{figure}

\begin{table}[bht]
\footnotesize
\centering
\caption{(Scenario 1) Mean of posterior mean and rescaled standard errors with coverage probabilities from DC and PDC. Estimates for DC are calculated excluding the local-trapped results.}
\label{cpcomparison}
\tabcolsep=8pt
\begin{tabular}{ccccccccc}
\hline
     & \multicolumn{2}{c}{DC ($k=12$)} & & \multicolumn{2}{c}{PDC ($k=12$)} & & \multicolumn{2}{c}{PDC ($k=30$)}\\
\cline{2-3} \cline{5-6} \cline{8-9} 
     Parameters  &  Estimate (SE) & CP & &  Estimate (SE) & CP & &  Estimate (SE) & CP\\    
\hline      
$\theta_1 = 2$ & 1.9957 (0.0101) & 52\% &  & 1.9971 (0.0101) & 92\% & & 1.9978 (0.0101) & 90\% \\
$\theta_2 = 1$ & 1.0060 (0.0102)& 48\% &  & 1.0045 (0.0102) & 82\% &  & 1.0037 (0.0101) & 82\% \\
$x_1(0) = 7$ & 6.9310 (0.1093) & 50\% &  & 6.9425 (0.1094) & 90\% &  & 6.9546 (0.1099) & 92\% \\
$x_2(0) = -10$ & -10.0869 (0.4279) & 58\% &  & -10.0908 (0.4254) & 86\% &  & -10.0921 (0.4244) & 88\% \\
$\sigma_1 = 1$ & 1.0026 (0.0646) & 60\% &  & 1.0057 (0.0647) & 96\% &  & 1.0094 (0.0650) & 96\% \\
$\sigma_2 = 3$ & 2.9737 (0.1917) & 58\% &  & 2.9681 (0.1919) & 92\% &  & 2.9545 (0.1920) & 92\% \\
\hline
\multicolumn{9}{l}{CP, Coverage Probability.}
\end{tabular}
\end{table}

Table \ref{cpcomparison} compares the two algorithms in terms of parameter estimation and coverage probabilities. The reported estimates and standard errors (SEs) are the means of posterior means and SEs computed across 50 different datasets.  The coverage probability for each parameter is calculated as the proportion of datasets in which the 95\% credible interval contains the true parameter value.  Since the DC method frequently gets trapped in local modes (demonstrated in Figure \ref{fig:local}), its coverage probabilities are significantly lower than those of PDC. Moreover, while PDC remains stable even with a larger $k=30$, DC fails to produce reliable posterior inference due to poor chain convergence.

\subsection{Scenario 2: Multiple True Parameters}

In the second simulation scenario, we consider a modified ODE model designed to produce a symmetric posterior with two true modes at $|\theta_1|=2$.  This scenario mimics situations where parameter estimates exhibit multimodality. We use the same priors and simulation settings as in Scenario 1, but generate data using the following ODE system:
\begin{align}\label{5.2equation}
    \frac{dx_1(t)}{dt} &= \frac{72}{(36+x_2(t))} -|\theta_1|, \nonumber\\
    \frac{dx_2(t)}{dt} &= \theta_2 x_1(t) - 1.
\end{align}
To evaluate how well each method captures the bimodal structure, we apply both DC and PDC to each synthetic dataset. 
\begin{figure}[htp]
    \centering
    \includegraphics[width=16cm, height=16.5cm]{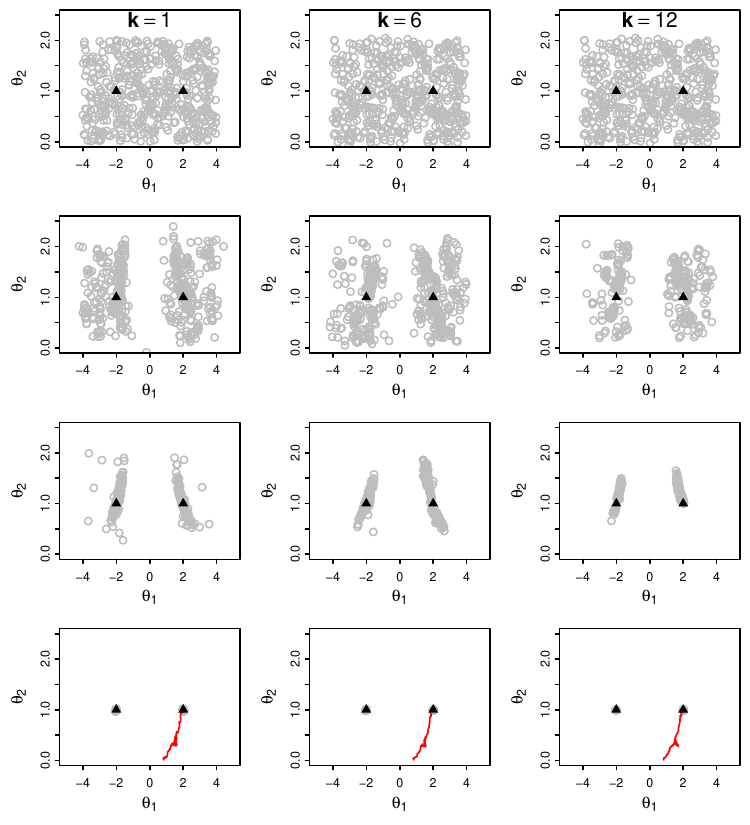}
    \caption{Intermediate posterior distributions for $\boldsymbol{\theta}$ by running PDC with three different $k=1, 6, 12$. The light gray empty circles are particles for $r=1, R_k/6, R_k/2, R_k$. The black triangular markers indicate true parameter values for generating ODEs. Solid lines on the 4th row are the trace plots from the standard DC for each $k$ value.}
    \label{fig:particles}
\end{figure}
Figure \ref{fig:particles} shows a graphical comparison of the two algorithms in estimating $\boldsymbol{\theta}$. Each column illustrates how particles for $\boldsymbol{\theta}$ gradually converge to the target distributions as the number of clones increases ($k=1,6,12$), while each row corresponds to the particles representing 
different stages of the intermediate posterior distribution. The light gray empty circles are particles at $r=1, R_k/6, R_k/2, R_k$, where $R_1 = 339, R_6=455, R_{12} = 511$. The last row in each column displays the particles from the final $k$-cloned posterior distribution, with the triangular markers denoting the true parameter values of $\boldsymbol{\theta}$. The solid lines in the last row represent the trace plots from the basic DC method for each $k$. The estimated parameters and corresponding SEs for both methods are summarized in Table \ref{tb:gss}. As shown, both DC and PDC yield posterior means of $\boldsymbol{\theta}$  close to the true values in Scenario 2. However, Figure \ref{fig:particles} illustrates that while PDC successfully captures both true modes simultaneously, the basic DC method's chains converge to only one mode, demonstrating the advantage of PDC in multimodal settings.

\begin{table}[hbt]
\footnotesize
\centering
\caption{(Scenario 2) Estimated parameters with standard errors.}
\label{tb:gss}
\tabcolsep=8pt
\begin{tabular}{cccccc}
\hline
     & \multicolumn{2}{c}{DC ($k=12$)} & & \multicolumn{2}{c}{PDC ($k=12$)} \\
\cline{2-3} \cline{5-6} 
     Parameters &  Estimate & SE & &  Estimate & SE \\    
\hline      
$|\theta_1| = 2$ & 2.0149 & 0.0098 &  & 2.0141 & 0.0109 \\
$\theta_2 = 1$ & 0.9946 & 0.0099 &  & 0.9953 & 0.0101 \\
$x_1(0) = 7$ & 7.1108 & 0.1000 &  & 7.1035 & 0.1025 \\
$x_2(0) = -10$ & -10.6346 & 0.4038 &  & -10.6332 & 0.3979 \\
$\sigma_1 = 1$ & 0.8920 & 0.0573 &  & 0.8927 & 0.0595 \\
$\sigma_2 = 3$ & 2.7491 & 0.1771 &  & 2.7488 & 0.1798 \\
\hline
\multicolumn{6}{l}{SE, Standard Error.}
\end{tabular}
\end{table}

\section{Real Data Analysis}
\label{sec:real}

\begin{figure}[htp]
    \centering
    \includegraphics[width=15cm]{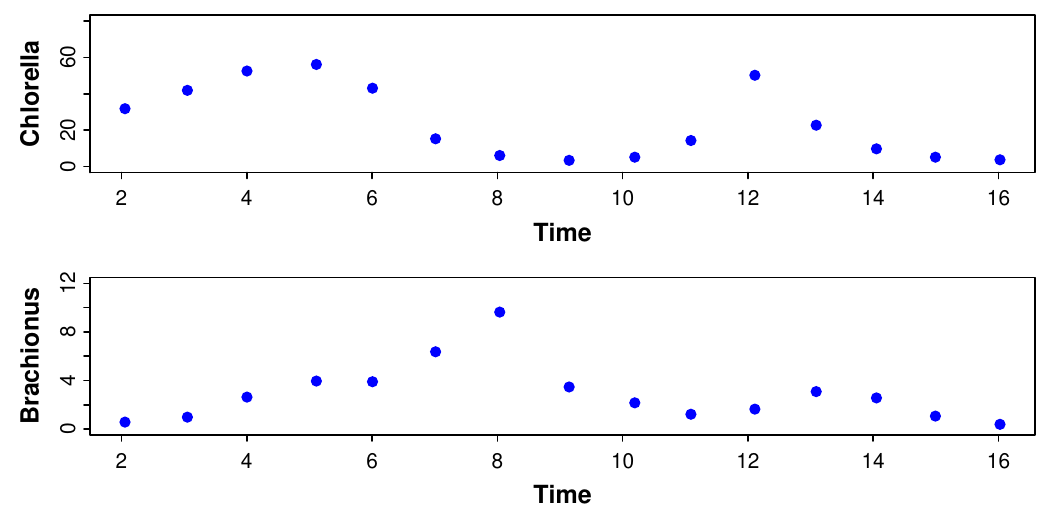}
    \caption{The experimental observation for the concentrations of \textit{Chlorella} and \textit{Brachionus} in the predator-prey dynamic system at the dilution rate $\delta = 0.68$ and the inflow nitrogen concentration $N^{\ast} = 80$.}
    \label{fig:prey68data}
\end{figure}
In this section, we conduct a real data analysis to demonstrate the benefits of PDC over the generalized profiling method, a popular frequentist approach for obtaining MLEs of ODE parameters. Specifically, we focus on the prey-predator ODE model described in \cite{fussmann2000}.  
In this nutrient-based predator-prey food cycle, planktonic rotifers (\textit{Brachionus calyciflorus}) consume unicellular green algae (\textit{Chlorella vulgaris}) whose growth is limited by the supply of nitrogen. The predator \textit{Brachionus} and the prey \textit{Chlorella} live alongside each other in repeated, experimental flow-through cultures known as chemostats. The system is theoretically modeled by \cite{fussmann2000} using a series of nonlinear ODEs linked by the interactions between the nitrogen resource, green algae, and planktonic rotifers as follows:
\begin{align}\label{5prey}
\frac{dN(t)}{dt} &= \delta(N^{\ast}-N(t))-F_C(N(t))C(t), \nonumber    \\
\frac{dC(t)}{dt} &= F_C(N(t))C(t)-F_B(C(t))B(t)/\epsilon-\delta C(t), \nonumber \\
\frac{dR(t)}{dt} &= F_B(C(t))R-(\delta+\alpha+m)R(t), \nonumber \\
\frac{dB(t)}{dt} &= F_B(C(t))R(t) - (\delta+m)B(t),
\end{align}
where $N(t), C(t), R(t), B(t)$ represent the concentrations of nitrogen, \textit{Chlorella}, reproducing \textit{Brachionus}, and total \textit{Brachionus}, respectively. The functional responses  $F_C(N(t)) = b_C N(t) / (k_C+N(t))$ and $F_B(C(t))=b_B C(t) / (k_B + C(t))$ include $b_C$ and $b_B$, the maximum birth rates of \textit{Chlorella} and \textit{Brachionus}, as well as $k_C$ and $k_B$, their respective half-saturation constants. The parameters $\epsilon, \alpha$, and $m$ represent the assimilation efficiency, fecundity decay,
and mortality of \textit{Brachionus}, respectively. All components of the dynamic system are extracted from the chemostats at a dilution rate $\delta$, while nitrogen 
 constantly enters the system at the same rate with an input concentration of $N^\ast$. 
\cite{Yosida2003} collected experimental data on the concentrations of \textit{Chlorella} and \textit{Brachionus} in the predator-prey dynamic system under an input nitrogen concentration $N^\ast = 80$ and a dilution rate $\delta=0.68$ (see Figure \ref{fig:prey68data}). 
Parameter estimation for the nonlinear dynamic system in Equation \ref{5prey} is particularly challenging since nitrogen ($N(t)$) and reproducing Brachionus ($R(t)$) concentrations are unobservable.

\begin{figure}[htp]
    \centering
    \includegraphics[width=15cm]{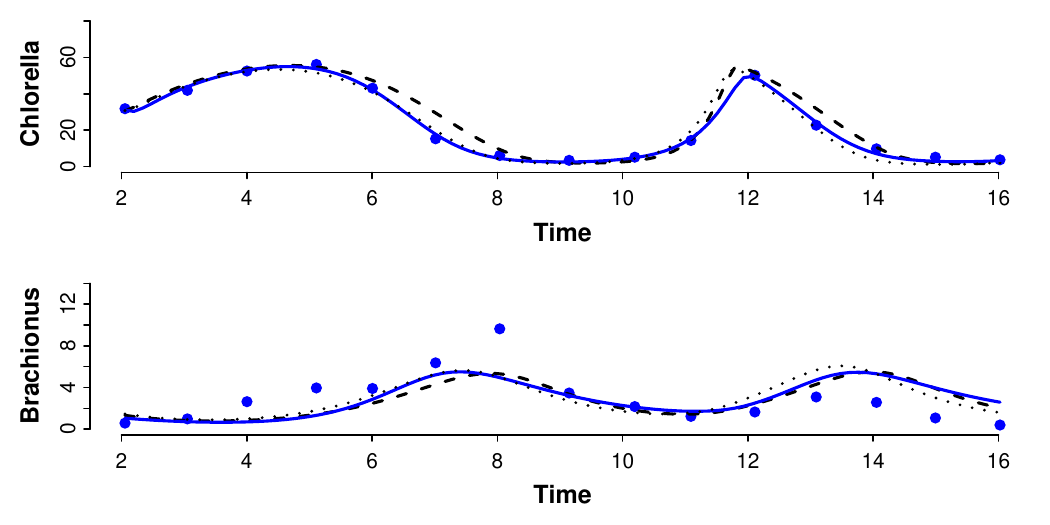}
    \caption{Solutions of the prey-predator ODEs in Equation \ref{5prey} using the parameter values from three methods: generalized profiling (dashed lines), PDC with $k=1$ (dotted lines), and PDC with $k=20$ (solid lines). Circles indicate the observed data points.}
    \label{fig:solutionsprey}
\end{figure}
 To implement particle data cloning, we set the rCESS threshold $\varepsilon_{\text{rCESS}} =0.999$ and the resampling threshold to $\zeta=0.5$. The total number of particles used throughout the algorithm is $M=500$. The prior distributions for the parameters of interest are based on the generalized profiling estimates from \cite{jiguorobust}, which serve as strongly informative priors in the Bayesian framework, particularly in cases with limited data. This prior setting allows us to assess whether PDC remains stable as a global optimizer,  even under strong prior knowledge. All priors are specified as normal distributions, except for $\sigma_1^2$ and $\sigma_2^2$, which follow an inverse gamma distribution, $IG(1,1)$. 

 We consider two candidate forward Markov kernels: $\pi_r$-invariant RWMH-Gibbs moves and $\pi_r$-invariant adaptive MH-Gibbs moves. The performance of DC and PDC under each kernel is compared in Table \ref{tb:kernelcomparison}. The maximum number of clones that each method can explore is denoted by $k_{max}$. We record the maximum log-likelihood ($Llik_{max}$) and the minimum rMSE ($rMSE_{min}$) until the posterior with $k_{max}$ is evaluated. Using the $\pi_r$-invariant RWMH-Gibbs kernel for both DC and PDC yields reasonable results; however, it fails to effectively explore larger values of $k$, making it insufficient in determining when to stop. In contrast, the $\pi_r$-invariant adaptive MH-Gibbs kernel allows PDC to stably explore the $k$-cloned posterior even for larger values of $k$ and demonstrates superior performance as a global optimizer. Meanwhile, the standard DC method fails to produce satisfactory results. Therefore, we select the $\pi_r$-invariant adaptive MH-Gibbs moves as the optimal kernel for running PDC. Based on this choice, we choose the number of clones to be $k=20$, as indicated by the standardized eigenvalue plot in Figure \ref{fig:ppeigen}.

 \begin{table}[ht]
\footnotesize
\centering
\caption{Comparison of DC and PDC for each forward kernel.}
\label{tb:kernelcomparison}
\tabcolsep=8pt
\begin{tabular}{cccccc}
\hline
    {Forward Kernels} & \multicolumn{1}{c}{$k_{max}$} & & \multicolumn{1}{c}{$Llik_{max}$} & & \multicolumn{1}{c}{$rMSE_{min}$} \\    
\hline      
$\text{DC}-\text{RWMH-Gibbs}$ & 15 & &  -80.344 & & 13.562   \\
$\text{PDC}-\text{RWMH-Gibbs}$ & 25 & &  -73.344 & & 11.288  \\
\hline
$\text{DC}-\text{Adaptive MH-Gibbs}$ & 20 & &  -73.605 & & 11.362  \\
$\text{PDC}-\text{Adaptive MH-Gibbs}$ & 50 & &  -54.426 & & 8.985 \\
\hline
\end{tabular}
\end{table}

Table \ref{tb:preyresults} shows parameter estimates obtained by applying our proposed method to the observed data with $\pi_r$-invariant adaptive MH-Gibbs moves.  PDC successfully produces valid results from each $k$-cloned posterior, allowing comparison  with the generalized profiling method in \cite{jiguorobust}. We also include the results of PDC with $k=1$, which corresponds to Bayesian inference using annealed SMC. Compared to the generalized profiling method,  particle data cloning with $k=20$ yields higher estimates for assimilation efficiency ($\epsilon$), decay of fecundity ($\alpha$), and the half-saturation constant of \textit{Brachionus} ($k_B$), while producing lower estimates for the mortality of \textit{Brachionus} ($m$), the half-saturation constant of \textit{Chlorella} ($k_C$), and the maximum birth rate of \textit{Chlorella} ($b_C$).

\begin{table}[ht]
\footnotesize
\centering
\caption{Estimated parameters of the predator-prey model and their standard errors using the generalized profiling method and PDC with $\pi_r$-invariant adaptive MH-Gibbs moves.}
\label{tb:preyresults}
\tabcolsep=8pt
\begin{tabular}{cccccc}
\hline
     & \multicolumn{1}{c}{Profiling} & & \multicolumn{1}{c}{PDC ($k=1$)} & & \multicolumn{1}{c}{PDC ($k=20$)} \\
\cline{2-2} \cline{4-4} \cline{6-6} 
      &  Estimate (SE) & &  Estimate (SE) & &  Estimate (SE)\\    
\hline      
$b_C$ & 3.9 (0.47) & & 3.586 (0.232) & & 3.501 (0.246)  \\
$b_B$ & 1.97 (0.26) & &  1.984 (0.129) & & 3.211 (0.301)  \\
$k_C$ & 4.3 (1.95) & &  4.331 (1.853) & & 0.007 (0.028)  \\
$k_B$ & 15.7 (2.01) & &  18.709 (1.732) & & 32.420 (4.938) \\
$\epsilon$ & 0.11 (0.02) & &  0.108 (0.011) & & 0.101 (0.014)  \\
$\alpha$ & 0.01 (0.14) & &  -0.089 (0.075) & & 0.249 (0.069) \\
$m$ & 0.152 (0.073) & &  0.124 (0.057) & & -0.042 (0.047)  \\
\hline
$rMSE$ & 20.157 & &  14.485 & & 9.965 \\
$Llik$ & -168.384 & &  -84.475 & & -65.322 \\
\hline
\multicolumn{6}{l}{SE: standard error.}
\end{tabular}
\end{table}

The solutions of the predator-prey ODEs in Equation \ref{5prey} are numerically obtained using parameter estimates from Table \ref{tb:preyresults}. Figure \ref{fig:solutionsprey} shows the resulting three ODE trajectories. To quantify goodness of fit, we compute the root mean squared errors (rMSE) between the solutions and the observations. Using particle data cloning using $k=20$, the rMSE is reduced by around 51\% compared to the generalized profiling method. Furthermore, our approach obtains a higher log-likelihood value, indicating better optimization of the complex likelihood surface of the predator-prey ODE model than the existing frequentist method. Since we employ a highly 
 informative prior based on \cite{jiguorobust}, the Bayesian inference result (PDC with $k=1$) is similar to that from the generalized profiling method. However, this result is influenced by strong prior assumptions and remains trapped in a local mode due to the limited number of data points. In contrast, our proposal with $k=20$ achieves a better fit, as reflected by lower rMSE and higher log-likelihood values. These results demonstrate that PDC can work well as a global optimizer, effectively overcoming local trapping and reliance on strong priors, even in data-scarce scenarios.

Table \ref{tb:adap2} compares the computation times for different initialization methods used in the PDC procedure, specifically our adaptive scheme and prior distributions. The computation times are measured while running PDC with varying numbers of clones in the predator-prey model discussed in Section \ref{sec:real}. Notably, the computation times for each $k$ value increase drastically with the prior-based initialization. In contrast, the computation times for the adaptive initialization method remain relatively stable. These results demonstrate that our adaptive scheme consistently outperforms the prior-based initialization in terms of computational efficiency.

\begin{table}[hbt]
\footnotesize
\centering
\caption{Comparison of computation times for PDC using different initialization methods (Adaptive vs. Prior) with varying numbers of clones in the prey-predator model.}
\label{tb:adap2}
\tabcolsep=8pt
\begin{tabular}{ccccccccc}
\hline 
     &  & $k=1$ & $k=5$ & $k=10$ & $k=20$ & $k=30$ & $k=40$ & $k=50$\\
\hline
     \multirow{ 2}{*}{Time (mins)} & Adaptive & 36.1& 32.2 &35.4&45.5&50.4&50.8&55.2 \\
     & Prior& 36.1 &44.3 &52.5&78.4&95.3&112.2&130.6\\
\hline     
\end{tabular}
\end{table}

\section{Discussion}
\label{sec:conc}
In this study, we propose particle data cloning (PDC) to estimate the MLEs of ODE parameters and their asymptotic variances. Traditional gradient-based frequentist methods often encounter the issue of local trapping when estimating unknown parameters from the multimodal likelihood surfaces of ODE models.  While standard MCMC-based data cloning (DC) theoretically mitigates multimodality, it frequently fails to explore the sharpened $k$-cloned posterior,  resulting in inaccurate parameter estimates. 

To address these challenges, we develop PDC to achieve global optimization for the ODE likelihood. PDC suppresses multimodality using cloned data while efficiently exploring the sharpened $k$-cloned posterior through intermediate distributions, facilitating flexible exploration of complex posterior spaces in the annealed SMC framework. We also design an adaptive initialization scheme to recycle information from the previous PDC results with smaller clone numbers. This scheme makes PDC more efficient in exploring the $k$-cloned posterior with larger values of $k$. Additionally, we provide practical guidelines for transforming MCMC kernels used in DC into their corresponding PDC versions.  

We validate the performance of PDC through two simulation studies, showing that our method estimates the ODE parameters more stably than MCMC-based data cloning in terms of coverage probabilities. Furthermore, our method successfully explores sharpened posterior distributions with a sufficiently large number of clones, leading to more reliable frequentist inferences.
In the second scenario of simulation studies, PDC accurately detects two true modes, whereas MCMC-based data cloning fails to do so. In the real-data application to a predator-prey model, PDC achieves a lower rMSE and a higher log-likelihood, highlighting its superior performance as a global optimizer compared to the generalized profiling method. Throughout the article, we emphasize that the uncertainty of estimates can be readily obtained from the particles, and the final results adhere to a frequentist interpretation. 

There are potential improvements and extensions of the proposed method for future research. One possible direction is to consider a more efficient algorithm to explore the target distribution with a large number of clones, $k$. Given that the performance of annealed SMC is contingent upon the Markov kernels used to rejuvenate particles, it is plausible to enhance the PDC approach by incorporating MCMC kernels with efficient proposal distributions. For example, it is possible to employ an adaptive SMC sampler \cite[][]{fearnhead2013}, which can automatically fine-tune the MCMC move kernels of an SMC sampler and select among various proposal densities, thereby obviating the need for manual selection of proposal distributions by the users. For another option, we can utilize the adaptive tuning of Hamiltonian Monte Carlo (HMC) within the SMC sampler as proposed by \cite{buchhol} to make the PDC efficient when model parameters have a large dimension. This modern approach combines HMC kernels with SMC samplers, outperforming simpler MCMC kernels such as random walk (RW) or Metropolis-adjusted Langevin (MALA) in high-dimensional settings. We can further study whether the proposed data cloning can be applied to any models with a prior for penalization. Especially, the semi-parametric Bayesian ODE model \cite[][]{Shijia2022} incorporates a model fidelity prior induced from \cite{ramsay2007parameter} while replacing $\boldsymbol{x}(t)$ in Equation \ref{ode_normal2} with penalized smoothing. It can take advantage of avoiding time-consuming numerical solvers for differential equations. We anticipate concentrating on these extensions in future research endeavors.


\section*{Acknowledgments}
The authors would like to thank Dr. Jiguo Cao from Simon Fraser University for providing the preprocessed real data used in their paper \citep{jiguorobust}, which enables us to conduct a fair comparison between our proposed method and theirs in Section \ref{sec:real}.

\nocite{*}
\bibliographystyle{asa}
\bibliography{sample}

\end{document}